\documentclass[aps,prb,twocolumn,showpacs]{revtex4-1}
\usepackage[latin1]{inputenc}
\usepackage{amsmath}
\usepackage{graphicx}
\usepackage{epstopdf}
\usepackage[LGRgreek]{mathastext}
\usepackage{graphicx}
\usepackage{grffile}
\usepackage{gensymb}
\usepackage{longtable}
\usepackage{lipsum}
\usepackage{array}

\begin{document}
\title{Crystal growth and magnetic anisotropy in the spin-chain ruthenate Na$_2$RuO$_4$} 
\author{Ashiwini Balodhi and Yogesh Singh}
\affiliation{Indian Institute of Science Education and Research (IISER) Mohali,
Knowledge City, Sector 81, Mohali 140306, India}

\begin{abstract}
We report single crystal growth, electrical resistivity $\rho$, anisotropic magnetic susceptibiltiy $\chi$, and heat capacity $C_p$ measurements on the one-dimensional spin-chain ruthenate Na$_2$RuO$_4$.  We observe variable range hopping (VRH) behaviour in $\rho(T)$. The magnetic susceptibility with magnetic field perpendicular ($\chi_\perp$) and parallel ($\chi_\parallel$) to the spin-chains is reported.  The magnetic properties are anisotropic with $\chi_\perp > \chi_\parallel$ in the temperature range of measurements T $\approx 2$ to $305$~K with $\chi_\perp / \chi_\parallel$ $\approx 1.4$ at $305$~K\@.  Analysis of the $\chi(T)$ data reveals an anisotropy in the $g$-factor and Van-Vleck paramagnetic contribution.  An anomaly in $\chi(T)$ and a corresponding lambda-like anomaly in $C_p$ at $T_N = 37$~K confirms long-range antiferromagnetic ordering.  This temperature is an order of magnitude smaller than the Weiss temperature $\theta \sim -250$~K and points to suppression of long range magnetic order due to low dimensionality.  However, we were unable to get a satisfactory fit of the experimental $\chi(T)$ by an isolated one-dimensional spin-chain model, suggesting the importance of inter-chain interactions in Na$_2$RuO$_4$. 
\end{abstract}
\maketitle
\section{Introduction}

Study of low dimensional magnets in the last few decades has led to the discovery of multiple quantum phases or systems.  The quasi one dimensional antiferromagnetic chain material Sr$_2$CuO$_3$ \cite{Schlappa, Motoyama}, Haldane-gap \cite{Haldane} in $S = 1$ spin chain compound Ni(C$_5$H$_{14}$N$_2$)$_2$N$_3$(PF$_6$)~ \cite{Honda}, spin-Peierls transition in CeCuGe$_3$ \cite{Hase1993}, realization of the Shastry-Sutherland model in SrCu$_2$(BO$_3$)$_2$, high-temperature superconductivity in cuprates, and quantum spin-liquid state (QSL) in triangular lattice organic compounds $\kappa$-(BEDT-TTF)$_2$Cu$_2$(CN)$_3$ \cite{Yamashita, Yamashita m} and in the Kagome bilayer magnet Ca$_{10}$Cr$_7$O$_{28}$~ \cite{Balz2016} etc., are just a few examples of the novel physics of low-dimensional magnets.  Enhanced quantum fluctuations due to reduced dimensionality in these materials provides a rich playground for the study of quantum phases.	

Recently, oxides with heavy transition metals ($4d, 5d$) have garnered much attention because of the possibility of novel magnetic behaviour arising from strong spin-orbit coupling \cite{Jackeli2009, Chaloupka2010, Pesin2010}.  The square lattice iridate Sr$_2$IrO$_4$ has been studied extensively for its structural and magnetic similarities with the parent high-$T_c$ cuprate material La$_2$CuO$_4$~\cite{Crawford1994,Kim2012, Korneta2010}.  The rare-earth pyrochlore iridate family $R_2$Ir$_2$O$_7$ has been studied for various novel behaviours like metal-insulator transitions \cite{Matsuhira2007} and possible topological properties \cite{Pesin2010, Yang2010, Wan2011, Zhang2017}.  Also, recently the two dimensional Kitaev candidate materials $A_2$IrO$_3$ ($A =$Na, Li) \cite{Chaloupka2010, Singh2010, Singh2012} and $\alpha-$RuCl$_3$ ~\cite{Plumb2014, Banerjee2016}, and the three-dimensional Kitaev materials $\gamma-$Li$_2$IrO$_3$ ~\cite{Modic2014} and $\beta-$Li$_2$IrO$_3$ ~\cite{Takayama2015} have been studied for their novel magnetic properties and possible proximity to Kitaev's quantum spin liquid phase.  Thus, there has been a resurgence in the interest in $4d-$ and $5d-$ based transition metal oxide (TMO) materials.   

Additionally, whether a novel $J_{eff} = 1/2$ localized model or a quasi-molecular orbital model is a more appropriate description for these compounds given their extended $d$-shells is also under debate.  The availability of single crystals of materials with heavy transition metal elements is therefore of importance for advanced measurements like X-ray absorption spectroscopy (XAS) or resonant inelastic X-ray scattering (RIXS) which help elucidate the electronic structure of the material.

In this work we focus on the $4d-$TMO linear spin chain ruthenate compound Na$_2$RuO$_4$ which combines low-dimensionality and strong spin-orbit coupling.  There have been some studies of the structural and magnetic properties of mostly polycrystalline Na$_2$RuO$_4$ \cite{Shikano, Mogre, Mogre m}.   
X-ray and neutron diffraction studies have shown that the structure of Na$_2$RuO$_4$ is quasi-one-dimensional with chains along the crystallographic $b$-axis built up of corner sharing RuO$_5$ trigonal bipyramids, where a Ru atom is surrounded by five oxygen atoms, making a linear one-dimensional spin-chain compound \cite{Mogre, Mogre m}.  Na$_2$RuO$_4$ has also been reported to show an antiferromagnetic transition around $T_N = 37$~K \cite{Shikano, Mogre, Mogre m}.  

In this work we report a crystal growth method to obtain relatively large single crystals of Na$_2$RuO$_4$.  We also report electrical transport along the chain direction, anisotropic magnetic susceptibility and magnetization, and heat capacity measurements on these crystals. 

\begin{figure}[ht]
\includegraphics[width=3cm]{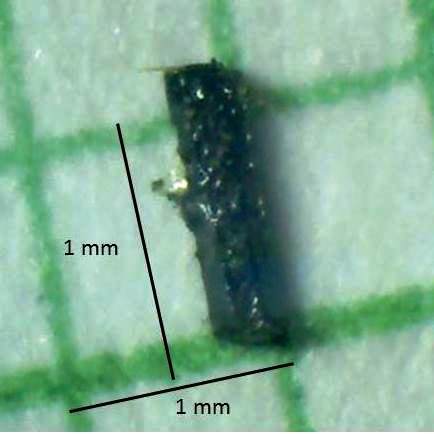}
\includegraphics[width=3cm, height = 3cm]{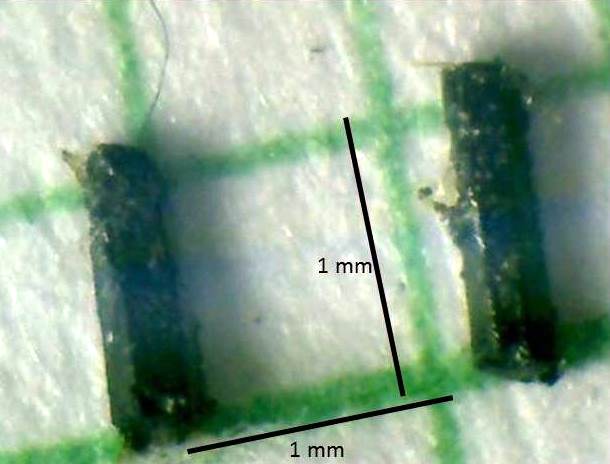}
\includegraphics[width=11 cm]{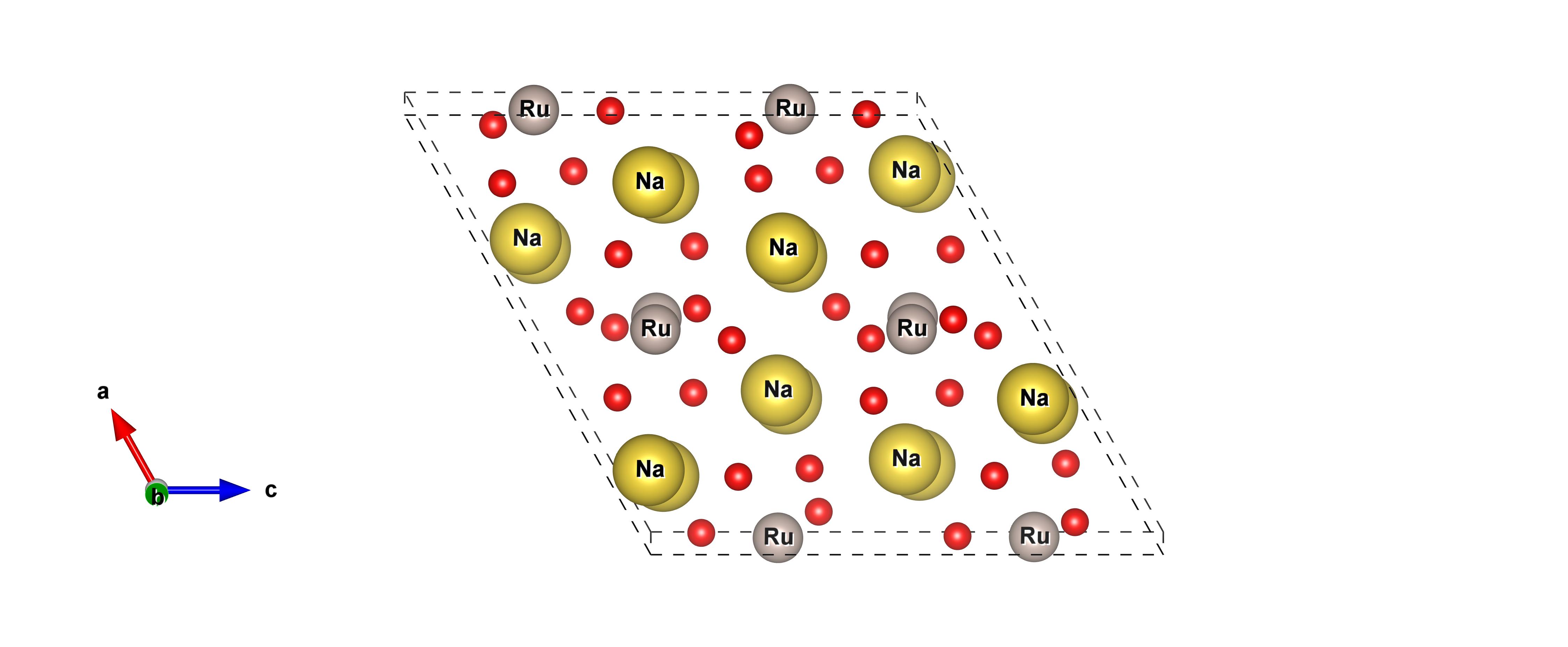}
\caption{(Top panel) Single crystals of Na$_2$RuO$_4$ on a millimeter grid.  (Bottom panel) The crystal structure of Na$_2$RuO$_4$ viewed down the chain direction (crystallographic $b$-axis).}
\label{Fig1}
\end{figure}

\section{Experimental details}
Single crystals of Na$_2$RuO$_4$ were grown from off-stoichiometric mixtures of Na$_2$O$_2$ and RuO$_2$ in an oxidizing atmosphere.  The high-purity starting materials Na$_2$O$_2$ ($93 \%$- Alfa-Aesar) and RuO$_2$ ($99.99 \%$- Alfa-Aesar) were taken in the ratio $1.5 : 1$ and mixed thoroughly inside an inert gas glove-box, pelletized, placed in an alumina crucible with a lid, and placed in a tube furnace.  The tube was evacuated and filled with oxygen.  The furnace was then heated to $750~^o$C in $5$~hrs and held there for $48$~hrs followed by slow cooling ($3$--$5~^o$C/hr) to room temperature.  Shiny needle like crystals with typical dimensions ($0.3 \times 1.3 \times 0.3)$ mm with the longest dimension being along the crystallographic $b$-axis, were obtained buried in a polycrystalline matrix.  Figure~\ref{Fig1} shows a few typical crystals obtained in this way.  The crystal structure was confirmed by single crystal X-ray diffraction on a Bruker diffractometer.   The measurement gave the space group P21/c and lattice parameters $a = 10.750$~\AA, $b = 7.036$~\AA, $c = 10.873$~\AA, and $\beta = 119.18^\circ$ which are in excellent agreement with previously reported values  \cite{Shikano, Mogre}. 
DC electrical transport, magnetic susceptibility, magnetization and heat capacity measurements were done using a Quantum Design physical property measurement system (QD-PPMS).

\section{Result and Discussion} 

\begin{figure}
\includegraphics[width=3.25 in]{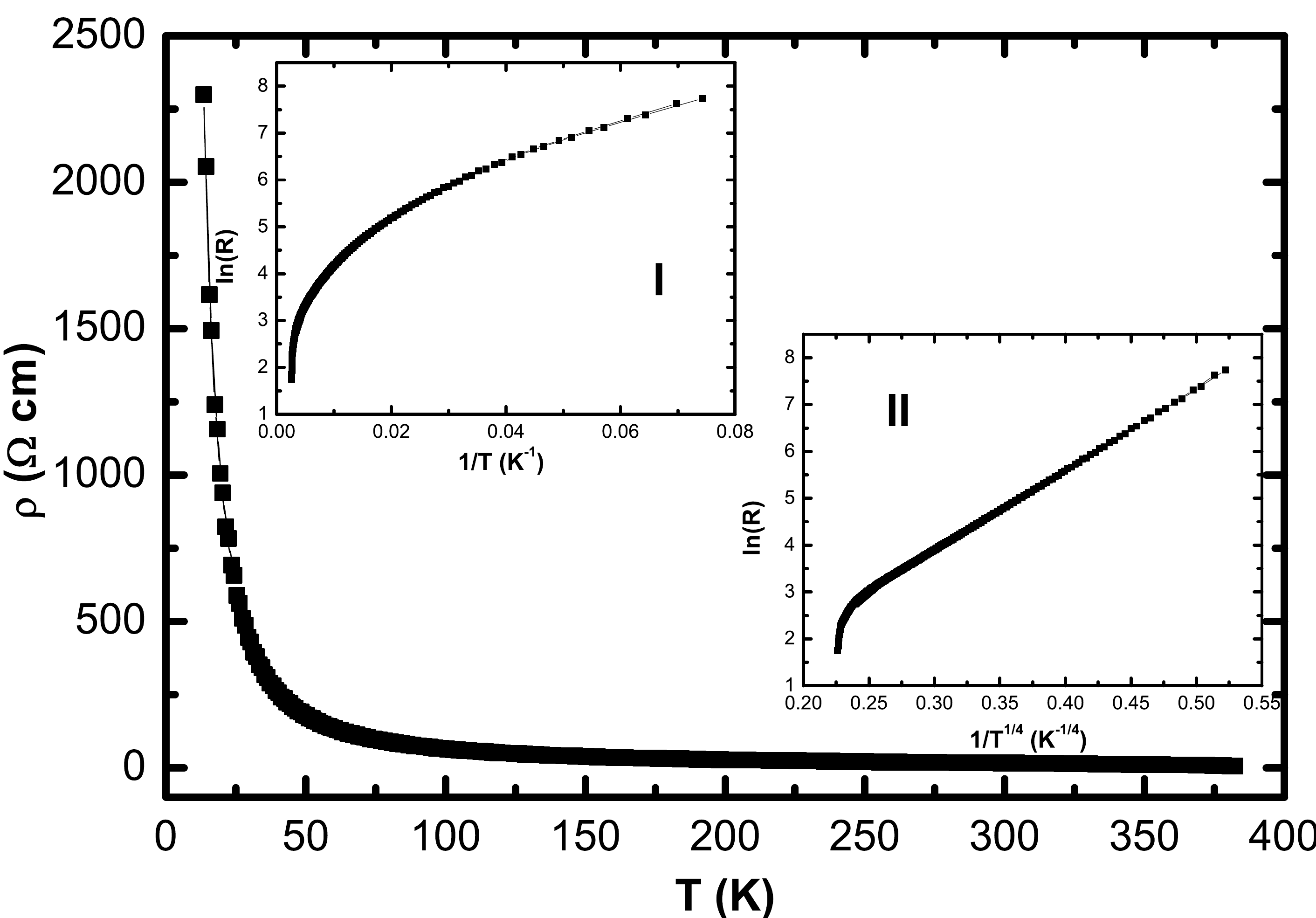}
\caption{ Electrical resistivity $\rho$ versus temperature $T$ for currents along the $b$-axis of Na$_2$RuO$_4$. Inset I shows a semi-log plot of $R$ vs $1/T$ data. Inset II shows a semi-log plot of $R$ vs $1/T^{1/4}$.} 
\label{fig:Fig_2}
\end{figure}

\begin{figure}
\includegraphics[width= 3.5 in]{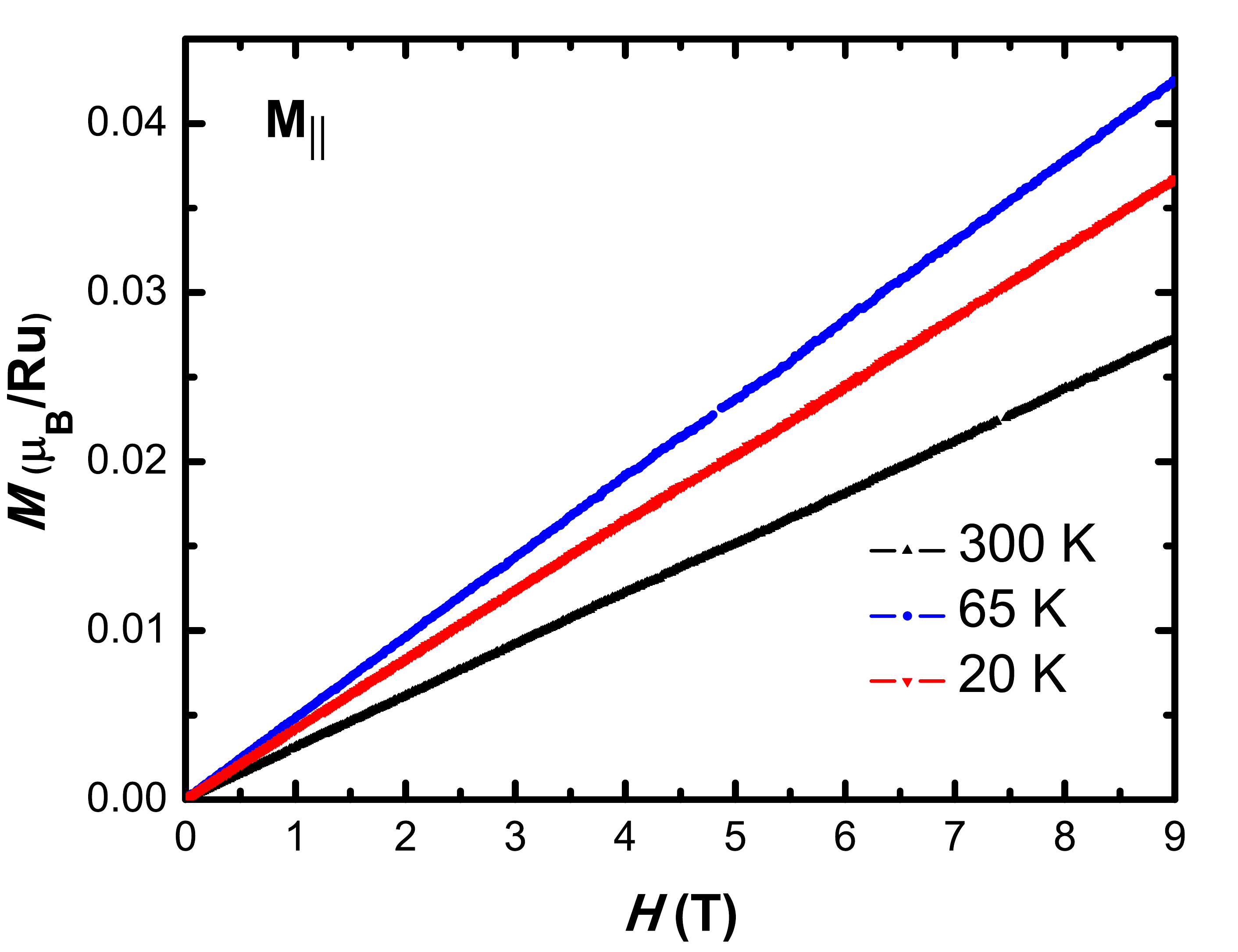}
\includegraphics[width= 3.5 in]{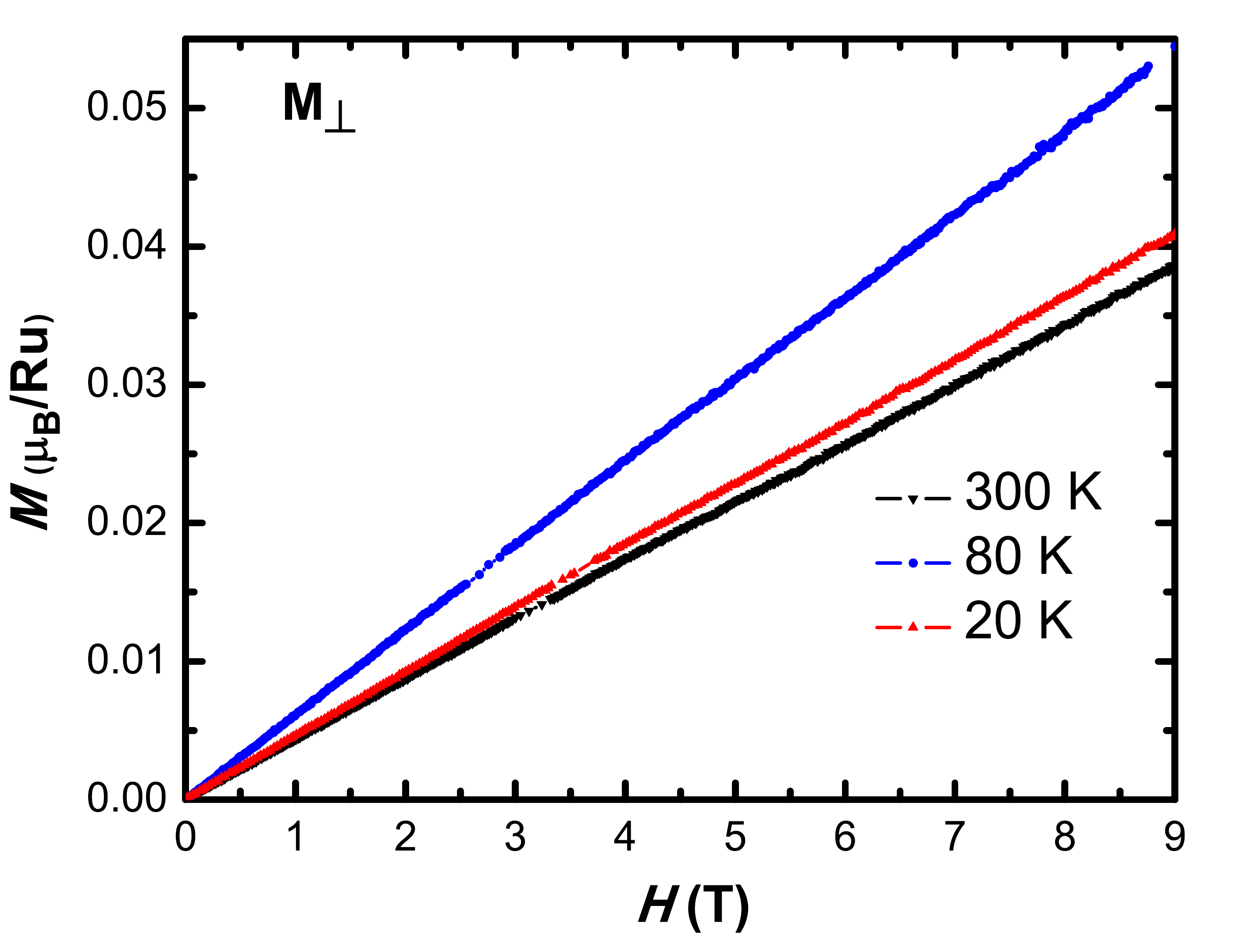}
\caption{ (color line) Magnetization $M$ versus magnetic field $H$ at various temperatures $T$ for Na$_2$RuO$_4$ single crystals with magnetic field $H$ parallel $M_{\parallel}$ (top panel) or perpendicular $M_{\perp}$ (lower panel) to the spin chain direction (crystallographic $b$-axis).}
\label{fig-MH}
\end{figure}
	
{\bf Electrical Transport:~~}Figure 2 shows the electrical resistivity $\rho$ versus temperature $T$ measured between $T = 5$~K and $T = 300$~K for a crystal of Na$_2$RuO$_4$ with a current $I = 1$~mA along the crystallographic $b$-axis. The $\rho(T)$ shows insulating/semiconducting behaviour in the whole temperature range. Inset I shows the data plotted as $R$ versus $1/T$ on a semi-log plot. From this plot it is clear that the data do not follow an Arrhenius kind of activated behaviour in any extended temperature range. Inset II shows the data plotted as $R$ versus $1/T^{1/4}$ on a semi-log plot.  Such a behaviour is expected when the conduction mechanism is variable range hopping (VRH) in three dimensions which is usually observed in disordered semi-conductors. It is clear that the $\rho(T)$ data for Na$_2$RuO$_4$ crystals have a temperature dependence which follows a VRH like behaviour over a large temperature range. The source of disorder in a high quality crystal is unclear at the moment. However, we point out that several transition metal oxides based on 4d- and 5d-transition metals have recently been shown to follow such transport behaviour. For example, transport on single crytals of Na$_2$IrO$_3$ ~\cite{Singh2010} and single crystals of Sr$_2$IrO$_4$ have also been observed to follow a VRH behaviour \cite{Rau}.  Therefore, a different common mechanism apart from disorder leading to this frequently observed behaviour in the transport of transition metal oxides can not be ruled out.\\

\begin{figure}
\includegraphics[width=3.25 in]{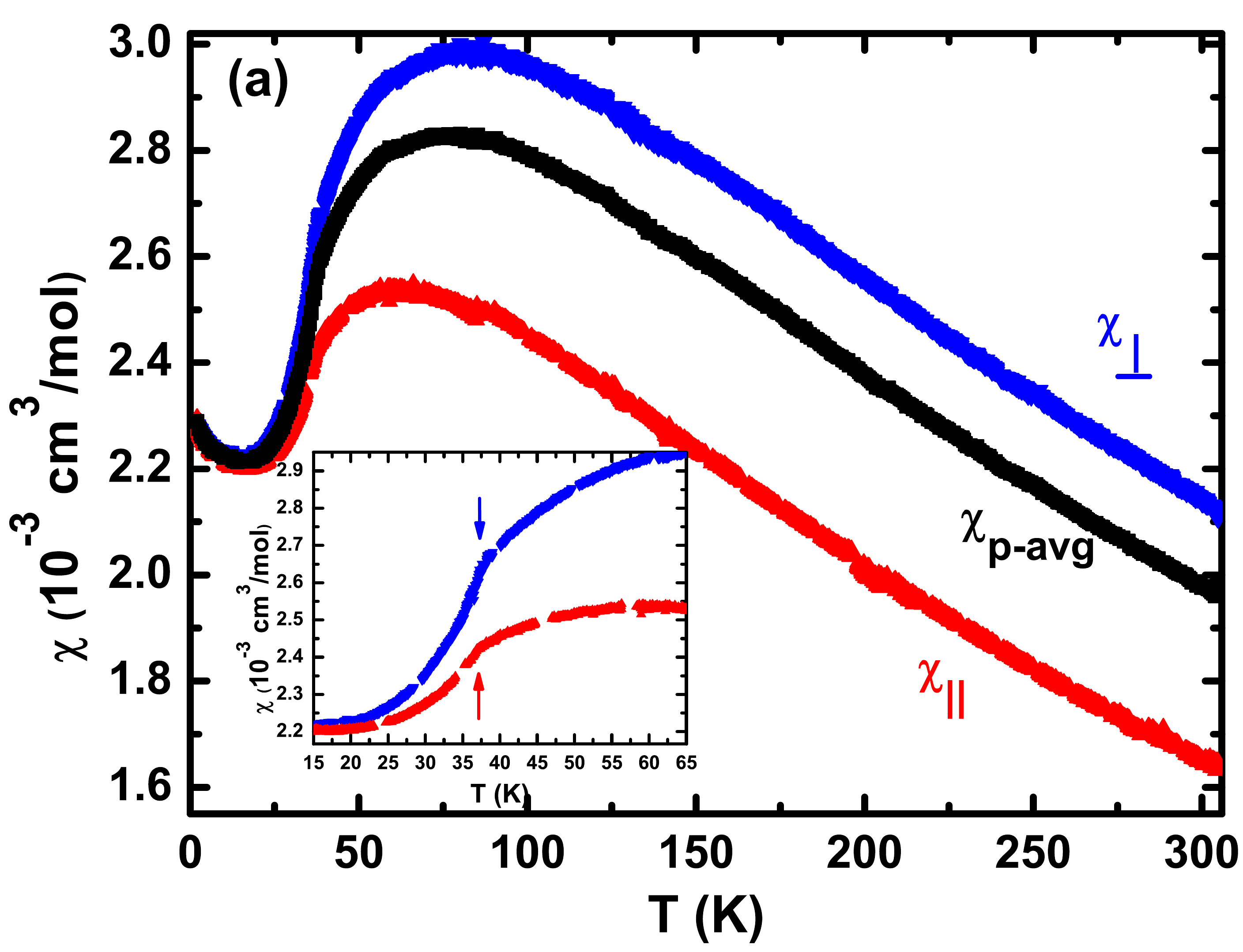}
\includegraphics[width=3.25 in]{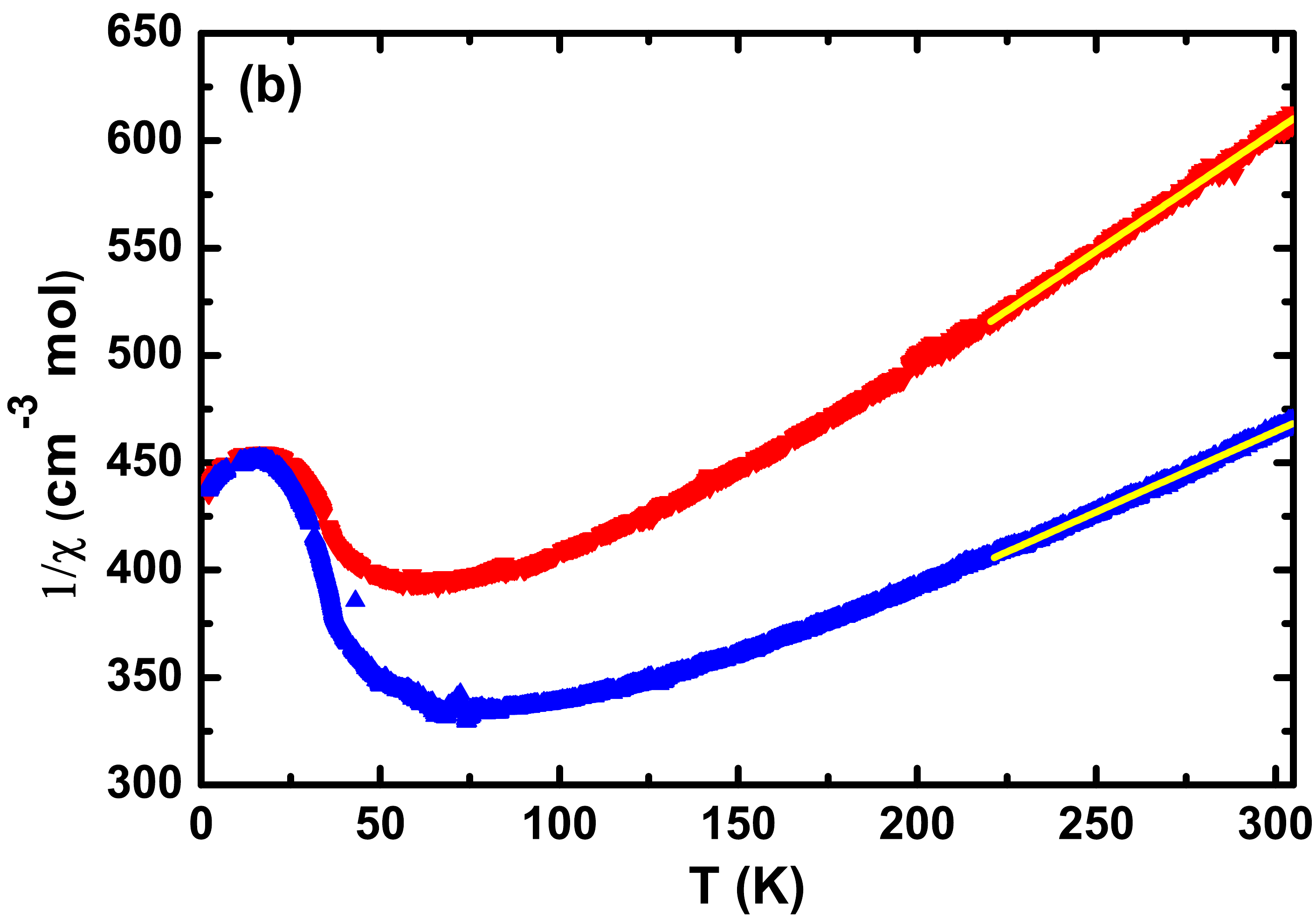}
\includegraphics[width=3.25 in]{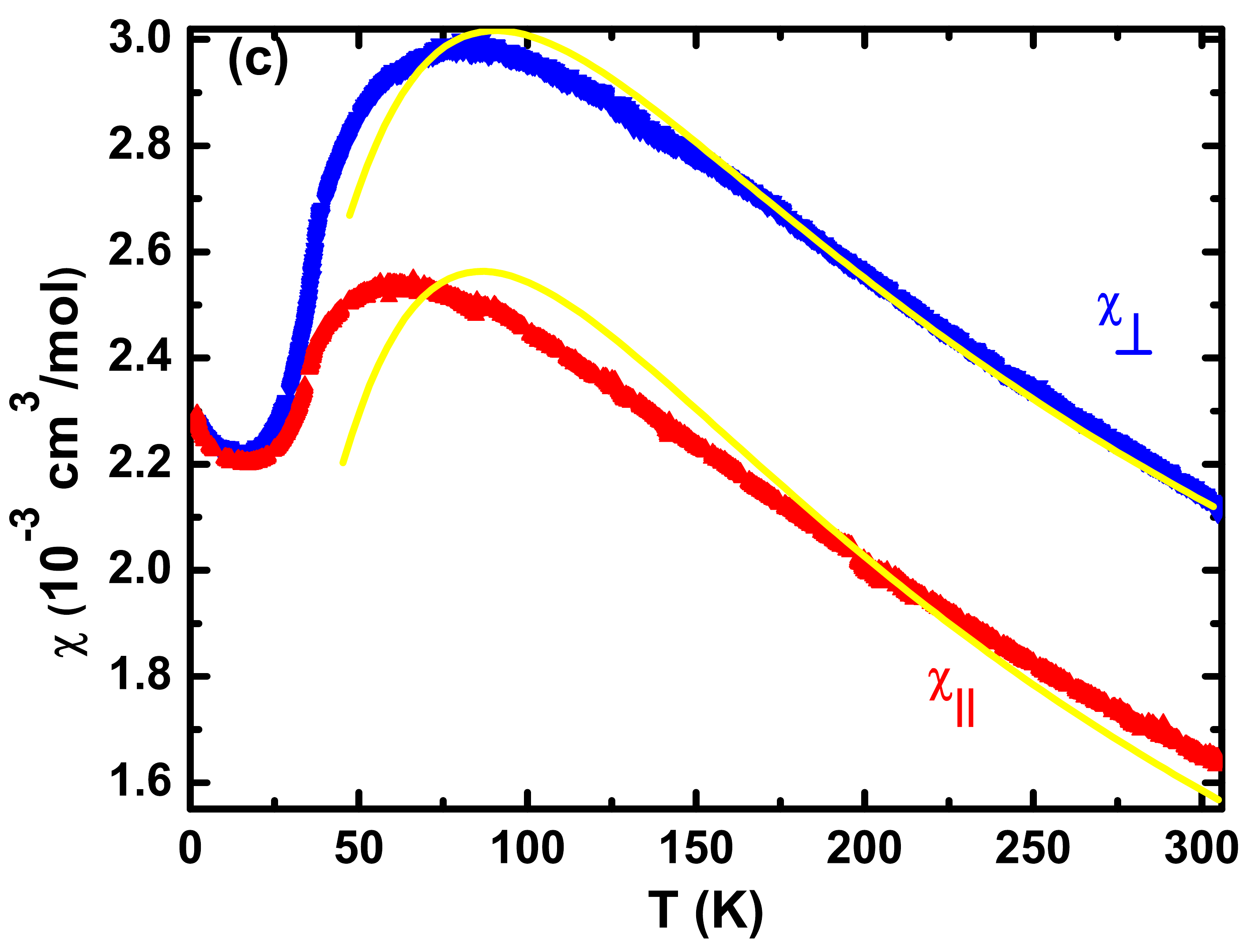}
\caption{ (color line) (a) Magnetic susceptibility $\chi$ versus temperature $T$ for Na$_2$RuO$_4$ single crystals with magnetic field $H = 1$~T applied parallel $\chi_{\parallel}$ or perpendicular $\chi_{\perp}$ to the spin chain direction (crystallographic $b$-axis).  Inset shows the $\chi(T)$ data below $T = 65$~K to highlight the antiferromagnetic transition marked by arrows at $T_N = 37$~K for both directions.  (b) $1/\chi(T)$ data for both field directions.  The solid curve through the data at high temperatures is a fit by the modified Curie-Weiss expression given in the text.  (c)  $\chi_{\parallel}$ and $\chi_{\perp}$ data along with fits (solid curves) to a one-dimensional spin chain model (see text for details). }
\label{fig-chi}
\end{figure}

{\bf Magnetization and Magnetic Susceptibility:~~} The magnetization $M$ versus magnetic field $H$ at different temperatures $T$ both above and below the magnetic ordering temperature $T_N$ is shown in Fig.~\ref{fig-MH}.   We observe that $M(H)$ is linear upto the highest magnetic fields $H = 9$~T\@.  The magnetic susceptibility $\chi = M/H$ measured with a magnetic field $H = 1$~T parallel ($\chi_{\parallel}$) and perpendicular ($\chi_{\perp}$) to the RuO$_5$ chains along the $b$-axis are shown in Fig.~\ref{fig-chi}~(a).  The powder average susceptibility defined for Na$_2$RuO$_4$ as $\chi_{\rm p-avg} = 2\chi_{\perp} + \chi_{\parallel}$ was also calculated and is plotted for comparison in Fig.~\ref{fig-chi}~(a).   The $\chi_{\rm p-avg}$ so obtained, is in good agreement with previous reports on polycrystalline samples  \cite{Shikano, Mogre, Mogre m}.  

From Fig.~\ref{fig-chi}~(a) the first thing to note is that $\chi_{\perp} > \chi_{\parallel}$ for all temperatures.  This anisotropy most likely occurs due to a $g$-factor and/or a Van-Vleck paramagnetic anisotropy in the material.   The broad maximum around $T\approx 74$~K observed for both directions is the behaviour typically observed for low-dimensional magnets and signals the onset of short range magnetic order.  At lower temperatures the $\chi$ decreases and there is a sharp cusp at $T_N = 37$~K observed for both directions signalling the long-range magnetic ordering of Ru$^{6+}$ ions. This can be seen in Fig.~\ref{fig-chi}~inset where the magnetic ordering for the two directions is indicated by arrows.  This is consistent with previously reported data on polycrystalline samples \cite{Shikano, Mogre m}.  A small upturn in magnetic susceptibility at the lowest temperatures suggests some Curie-like paramagnetic contribution most likely arising from impurities or unpaired spins.

\begin{table}[b]
\caption{ Parameters obtained by fitting anisotropic magnetic susceptibility of Na$_2$RuO$_4$}
\begin{tabular}  {|c|c|c|c|c|}	
\hline	
Alignment  & $\chi_0$~(cm$^3$/mol) &$\theta$~(K)  \\ \hline 
$\chi_{\parallel}$    & $1.08(2) \times 10^{-5}$&$-271(1)$ \\ \hline 
$\chi_{\perp}$  & $3.03(3) \times 10^{-4}$ &$-242(1)$ \\			
\hline
\end{tabular}
\label{Table-fit}
\end{table} 

From the $\chi(T)$ data we see that for both field directions as we cool from $T = 300$~K the data shows an upward curvature and below about $T = 150$~K the data change behaviour and curves downwards.  Therefore, for $T \geq 200$~K we believe that we are in the high $T$ paramagnetic regime where a Curie-Weiss analysis can be used.  The $\chi(T)$ data above $T = 220$~K for both directions were fit by a modified Curie-Weiss expression $\chi = \chi_0 + C/(T-\theta)$, with the temperature independent contribution $\chi_0$, the Curie constant $C$, and Weiss temperature $\theta$ as fit parameters.  It was found that the parameters $\chi_0$, which arises from a combination of core diamagnetism (isotropic) and Van-Vleck paramagnetism (anisotropic), and $C$ which will be anisotropic due to $g$-factor anisotropy, could not be varied simultaneously to get unique values for these two fit parameters.   In the fits we therefore fixed $C$ to the value expected for $S = 1$ with a $g$-factor $g = 2$.  The fits for both field directions are shown as the solid curve through the $1/\chi(T)$ data shown in Fig.~\ref{fig-chi}~(b) and the values of $\chi_0$ and $\theta$ obtained from the fits are given in Table~\ref{Table-fit}.  These results show that the Van Vleck paramagnetic contribution is an order of magnitude different for the two orientations.  Since anisotropy in the Van Vleck contribution is linked to a $g$-factor anisotropy, it is most likely true that the $g$-factor itself is highly anisotropic between the two directions.  We however, cannot find the anisotropy in both quantities simultaneously from our fits.  

The Weiss temperatures $\theta$ were found to be similar for $\chi_{\parallel}$ and $\chi_{\perp}$, respectively.  These $\theta$ values are large and negative indicating strong antiferromagnetic exchange interactions between the $S = 1$ moments.  The observed magnetic ordering occurs at $T_N = 37$~K, which is an order of magnitude smaller than $|\theta|$ indicating strong low-dimensionality which suppresses the long ranged ordering to much lower temperatures.  To further explore the effect of low-dimensionality on the magnetism, we have attempted to fit our $\chi(T)$ data by a phenomenological expression for the magnetic susceptibility of the $S = 1$ Haldane spin-chain \cite{Souletie2004}: $$\chi(T) = \chi_0 + {0.125\over T}exp(-{0.451 J\over T})+{0.564\over T}exp(-{1.793 J\over T}),$$ where $\chi_0$ is a $T$ independent term and $J$ is the magnitude of the exchange interactions betwen the $S = 1$ spins within the chains.  Fits to $\chi_{\parallel}$ and $\chi_{\perp}$ data for $T \geq 50$~K, were performed using the above expression and the best fits, shown as the solid curves through the data in Fig.~\ref{fig-chi}~(c), gave the value $J \approx 72$~K\@.  It is evident that while the fit reproduces qualitatively the main features (high $T$ Curie like behaviour and a broad maximum) of the data, the quantitative match is not very good.  This suggests that substantial inter-chain couplings need to be included in any modelling of the data and that a model of isolated chains is not sufficient to understand the magnetism in Na$_2$RuO$_4$.  The presence of substantial inter-chain couplings is already evidenced by the presence of long range magnetic order below $T_N = 37$~K\@.  \\  

\begin{figure}
	\centering
	\includegraphics[width=8.5cm, height=6cm]{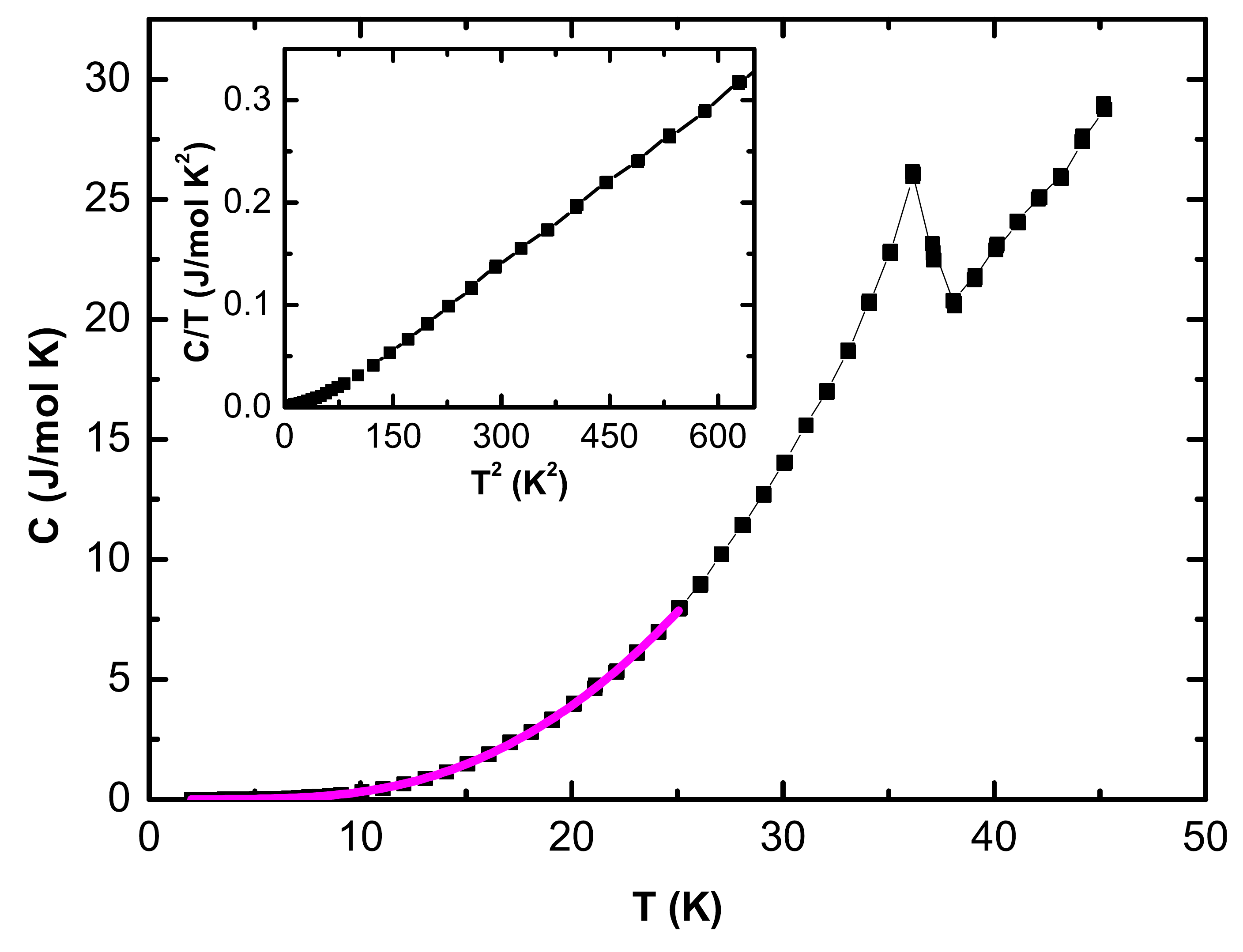}
	\caption{ Heat Capacity C$_p$ versus temperature for Na$_2$RuO$_4$ between T = 2 to 50 K.}
\label{fig-Cp}
\end{figure}

{\bf Heat Capacity:~~}The heat capacity data in zero magnetic field is plotted in Fig.~\ref{fig-Cp}.  The bulk nature of the long-ranged magnetic ordering in Na$_2$RuO$_4$ is confirmed by a sharp $\lambda$-type anomaly at $T = 37$~K consistent with magnetic susceptibility data presented above.  The unavailability of an iso-structural non-magnetic material which can be used as an approximate lattice contribution to the heat capacity for Na$_2$RuO$_4$ makes an analysis of magnetic entropy released at $T_N$ difficult.  Heat capacity data below the ordering temperature can give information about the magnetic excitations.  We attempted to fit the $C_p(T)$ data below $T_N$ by various models.  We tried a fit to the data below $T = 20$~K by the expression $C_p = \beta_1 T + \beta_2 T^3$.  Since Na$_2$RuO$_4$ is an insulator, the origin of the linear-$T$ term in the heat capacity comes from the contribution of one-dimensional antiferromagnetic magnons.  The $T^3$ term is the usual contribution from phonons plus possible contributions from three dimensional antiferromagnetic magnons.  The above expression gave a very poor fit to the data and the best fit gave a negative value for the prefactor of the linear-$T$ term which would be unphysical.  
Figure~\ref{fig-Cp}~inset shows the $C_p/T$ versus $T^2$ data inside the magnetically ordered state.  It can be seen from this plot that the data at the lowest temperatures show a departure from the conventional $C_p \sim T^3$ behaviour and falls more rapidly.  We therefore attempted and were successful in getting an excellent fit to the following model with $C_p(T) = \beta T^3 + AT^{3/2}exp^{-\Delta / T}$, where the second contribution is from a possible spin-gap (the exponential) and the $T^{3/2}$ pre-factor is from excitations of the ferromagnetically coupled spins within the chains.  The fit is shown as the solid curve through the heat capacity data below $T = 25$~K in Fig.~\ref{fig-Cp}.  \\

\section{Conclusion}

In summary, we have presented a crystal growth method for obtaining sizeable single crystals of the spin-chain ruthenate Na$_2$RuO$_4$ big enough for transport and anisotropic magnetic measurements.  We have measured electrical transport with current along the chain direction, anisotropy in the magnetic susceptibility, and heat capacity on these crystals.  The  electrical transport along the crystallographic $b$-axis shows a variable range hopping behaviour over a large temperature range.  Although VRH behaviour is usually associated with disorder and is observed in some doped semiconductors, we note that several $4d$-- and $5d$-- TMO's including Sr$_2$IrO$_4$ and Na$_2$IrO$_3$ have been reported to show such behaviour.  Anisotropy is clearly observed in magnetic measurements with $\chi_{\perp} > \chi _{\parallel}$.  From an analysis of the $\chi(T)$ data we conclude that this anisotropy arises from a combination of anisotropy in the $g$-factors as well as in the Van-Vleck paramagnetic contribution.  The Weiss temperatures $\theta_{\parallel} = -271$~K and $\theta_{\perp} = -242$~K for the two field directions are large and antiferomagnetic.   Both magnetic and heat capacity measurements confirm long-ranged antiferomagnetic order below $T_N = 37$~K\@.  A suppression of $T_N$ compared to $\theta$ by almost an order of magnitude most likely points to low-dimensionality.  However, we were unable to obtain satisfactory fits to our $\chi(T)$ data to a model of isolated $S = 1$ spin-chains.  This suggests the importance of inter-chain exchange interactions in Na$_2$RuO$_4$.  Heat capacity data in the magnetically ordered state could also not be fit using a model of one-dimensional magnons and phonons.  This is somewhat surprising given that Na$_2$RuO$_4$ is a fairly good example of a quasi-one-dimensional spin-chain magnet with large intra-chain interactions but a much smaller ordering temperature.  We were able to obtain good fits to the $C_p(T)$ data below $T_N$ by a model with a spin excitation gap.  

Future measurements like electron spin resonance (ESR) and nuclear magnetic resonance (NMR) would be useful to get accurate estimates of the anisotropic $g$-factor and the Van-Vleck term in the susceptibility.

\end{document}